\documentclass[a4paper,11pt]{article}
\usepackage{pos}
\usepackage{multirow}
\usepackage{upgreek}
\usepackage{subcaption}
\usepackage{wrapfig}
\usepackage{enumitem}
\setenumerate{itemsep=0mm}

\title{Cosmic-Ray Studies with the Surface Instrumentation of IceCube}
\ShortTitle{Cosmic-ray studies at IceCube}

\author{Andreas Haungs for the IceCube Collaboration}

\affiliation[a]{Karlsruhe Institute of Technology, Institute for Astroparticle Physics, Germany}

\emailAdd{andreas.haungs@kit.edu}

\abstract{IceCube is a cubic-kilometer Cherenkov detector installed in deep ice at the geographic South Pole. IceCube's surface array, IceTop, measures the electromagnetic signal and mainly low-energy muons from extensive air showers above several 100 TeV primary energy, with shower bundles and high-energy muons detected by the in-ice detector IceCube. In combination, the in-ice detector and IceTop provide unique opportunities to study cosmic rays in detail with large statistics. This contribution summarizes recent results from these studies. In addition, the IceCube-Upgrade will include a considerable enhancement of the surface detector through the installation of scintillation detectors and radio antennas and possibly small air-Cherenkov telescopes. We will discuss the results of the prototype detectors installed at the South Pole and the prospects of this enhancement as well as the surface array planned for IceCube-Gen2. 

\vspace{4mm}
{\bfseries Corresponding author and presenter:}
Andreas Haungs \\[4mm]

\FullConference{37$^{\rm{th}}$ International Cosmic Ray Conference (ICRC 2021)\\
		July 12th -- 23rd, 2021\\
		Online -- Berlin, Germany}

}

\begin{document}
\maketitle

\section{Cosmic-Ray Physics at the South Pole}
\label{sec:info}

Experimental cosmic-ray (CR) research aims to determine the energy spectrum, the elemental composition, and the arrival direction distribution of incoming cosmic particles. Such measurements are essential for understanding the sources, acceleration, and propagation of these energetic particles of cosmic origin. 
At energies above $10^{14}\,$eV, the characteristics of these particles are mostly determined indirectly from measured properties of the extensive air showers (EAS) induced by primary cosmic rays in Earth's atmosphere.

%
\begin{wrapfigure}{l}{0.45\textwidth}
  \begin{center}
    \includegraphics[width=0.43\textwidth]{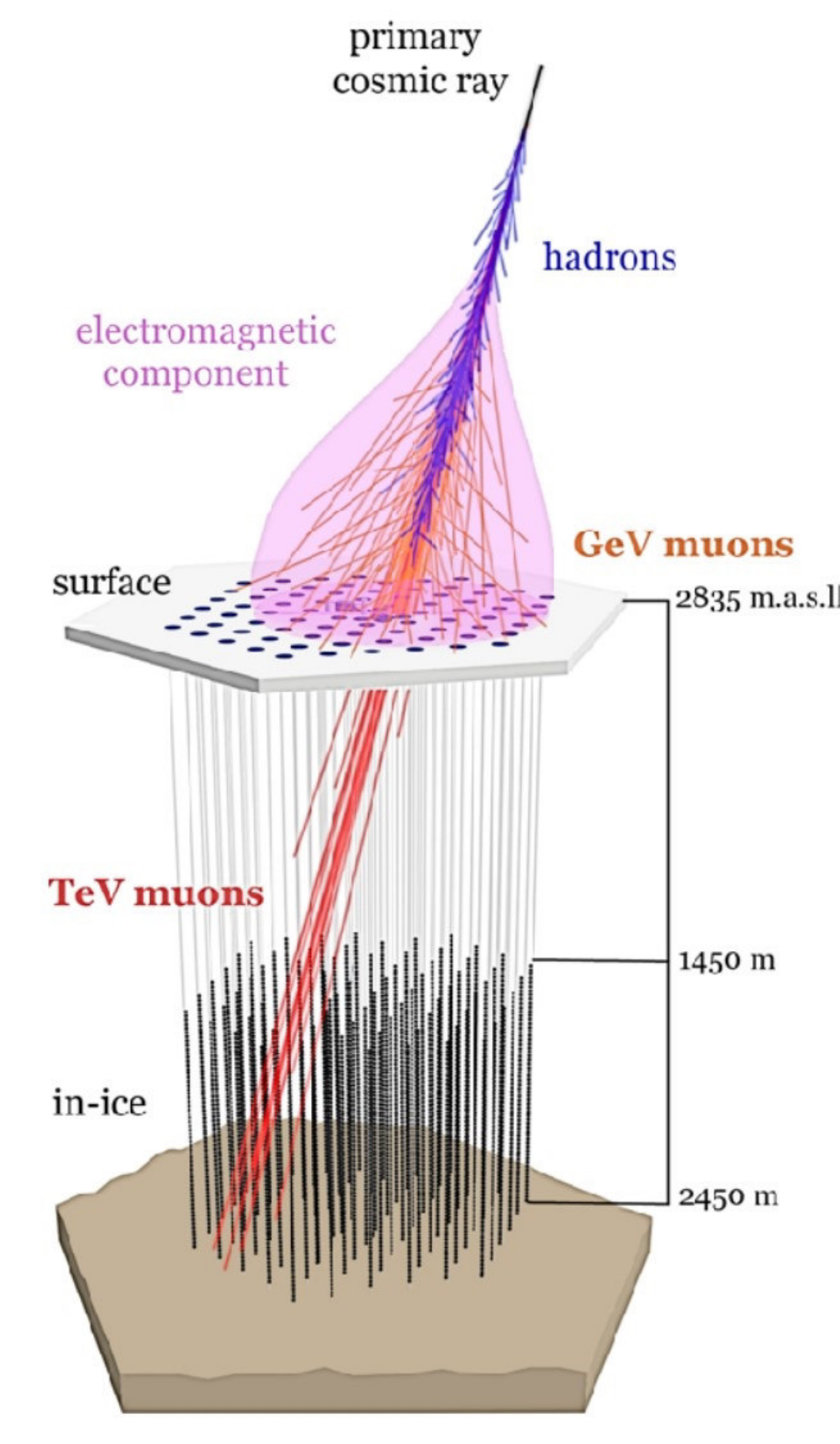}
  \end{center}
  \caption{Scheme of air-shower measurements at the IceCube Neutrino Observatory~\cite{Verpoest:2021icrc}}
  \label{fig1}
\end{wrapfigure}
IceCube's surface array, IceTop~\cite{IceCube:2012nn}, has proven over the past decade to be a very valuable detector component not only for the calibration of IceCube, but also in combination with the unique cube-kilometer-sized in-ice muon detector as a cosmic ray observatory (Fig.~\ref{fig1}). 
It is providing veto and calibration functionality for the in-ice neutrino measurements~\cite{Tosi:2019nau}, as well as measurements of the primary cosmic-ray spectrum  and mass-composition from 1 PeV to about 1 EeV~\cite{IceCube:2019hmk}. 
The latter is -- beside knowledge gain in the PeV to EeV primary energy range of cosmic rays -- essential to reduce systematic uncertainties on the atmospheric backgrounds of astrophysical neutrinos in the ice~\cite{Gaisser:2016obt}.
Any progress in the research field depends on the validity of hadronic interaction models required for the interpretation of EAS measurements. Hence, it is important to reduce the uncertainties by improving hadronic interaction models and enhancing air-shower arrays to perform hybrid measurements of the various EAS components.
Moreover, IceTop contributes to IceCube's multi-messenger mission in particular regarding Galactic sources: searches for photons~\cite{Aartsen:2012gka}, and measurements of the anisotropy of Galactic cosmic rays~\cite{IceCube:2013mar}.

IceCube and other experiments have contributed to today's knowledge of Galactic cosmic rays (GCR) which can be summarised as follows.
The all-particle spectrum has a steep power-law like behaviour with features known as `knee', 'second knee' and `ankle' at $2$-$5\cdot10^{15}\,$eV, $1$-$3\cdot10^{17}\,$eV and $2$-$8\cdot10^{18}\,$eV, respectively. 
Whereas at the knee and the second knee the spectrum steepens, the ankle is characterised by a flattening of the spectrum.
Cosmic rays below the knee are of galactic origin and cosmic rays above the ankle are most probably of extra-galactic origin. 
Somewhere in the energy range from $10^{16}\,$eV to a few $10^{18}\,$eV the transition of cosmic rays from galactic to extra-galactic origin is expected. 
There are, however, still major issues regarding the highest energy GCR:
(i) The most powerful accelerators of cosmic rays in our Milky Way have not yet been revealed. 
(ii) The maximum energies of various possible acceleration mechanisms and sources are uncertain.
(iii) The Galactic extra-galactic transition and features in the CR energy spectrum
are not well understood. 
These questions can be addressed through improved measurements of the energy dependent composition of GCR in conjunction with gamma-ray and neutrino observations. 
It means that we must bring multi-messenger astrophysics to maturity not only at the ultra-high energy range, but also for the Galactic scenario at lower energies. 

In this contribution we summarise the recent achievements and future plans for air-shower measurements by the IceCube collaboration, which are described in more detail in further papers for this conference.

\section{Recent Results}
\label{sec:recent}

\begin{wrapfigure}{r}{0.55\textwidth}
  \begin{center}
    \includegraphics[width=0.53\textwidth]{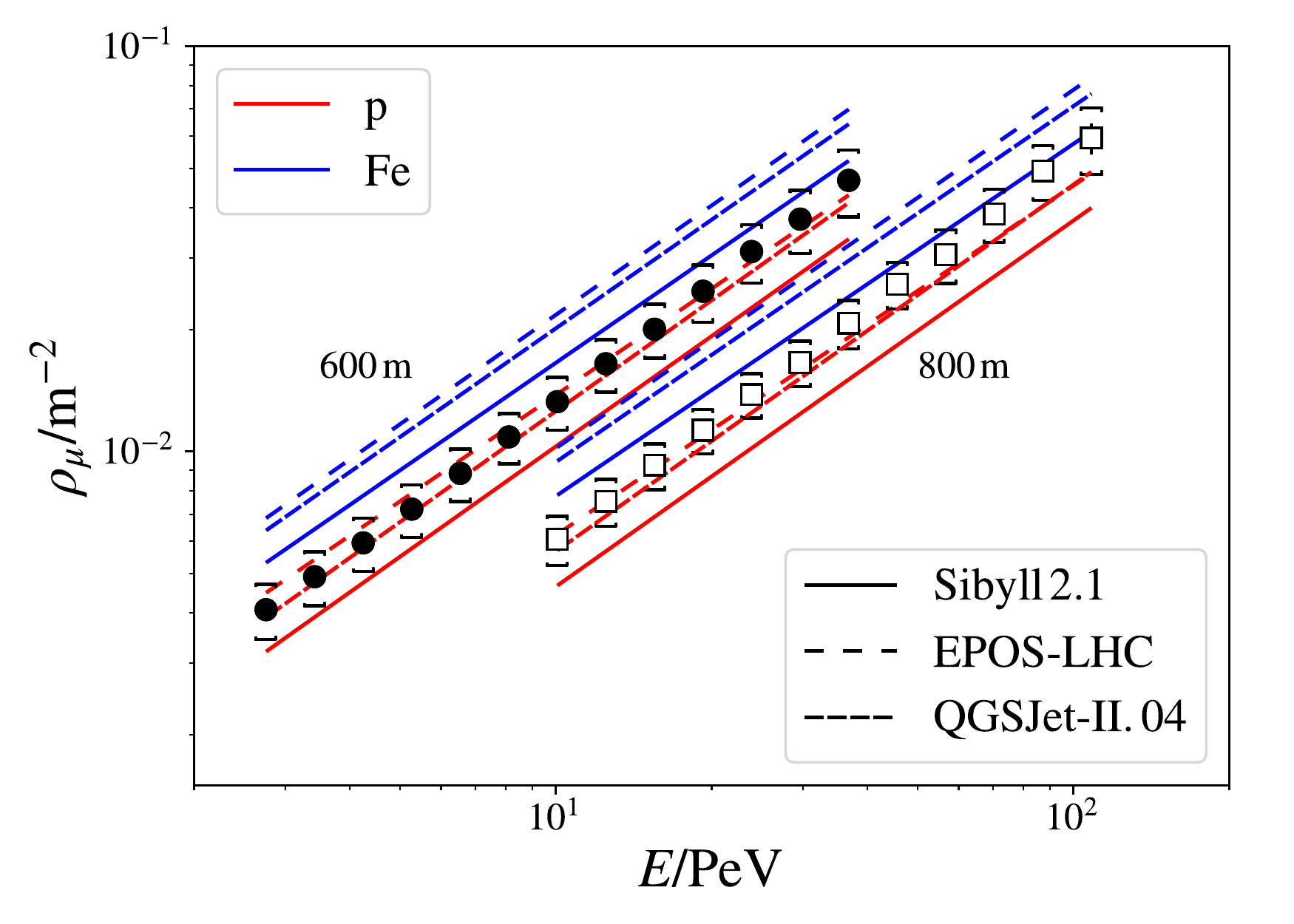}
  \end{center}
  \caption{Measured muon density at 600 m (circles) and 600 m (squares) lateral distance. Error bars indicate the statistical, brackets the systematic uncertainty. Shown for comparison are the corresponding simulated densities for proton and iron (red and blue lines)~\cite{Soldin:2021icrc}.}
  \label{fig2}
\end{wrapfigure}
Important for understanding GCR is first and foremost a precise knowledge of the all-particle energy spectrum and the elemental composition of cosmic rays in the entire transition range from 100 TeV to 10 EeV of primary energy. 
IceCube has delivered important milestones, first with the composition paper~\cite{IceCube:2019hmk} from three years of data taking and now recently with the reconstruction of the lower energy spectrum in the 100 TeV - 1 PeV range~\cite{IceCube:2020yct}. 
With a new trigger that selects events in closely spaced detectors in the center of the array, the IceTop energy threshold has been lowered by almost an order of magnitude below its previous threshold of 2 PeV. New machine-learning methods were developed to deal with events with very few detectors hit. In Figure~\ref{fig:spectrum} the results are compared with previous measurements by IceTop and other experiments. 
However, the results also show that there is still a long way to go to understand high-energy GCR. 
In particular, both the statistical and systematic uncertainties need to be reduced, and the analyses need to be extended to a coherent determination of the elemental composition and possible small and large scale anisotropies in a wide energy range. In addition, the understanding of shower development needs to be further improved.
For this, IceCube offers the best prerequisites: a total of 9 years of data are now available, new (machine learning) methods have been developed and are being applied, and with new detector components the phase space as well as the accessible energy ranges will be extended to higher and lower energies in the future. The following chapters give an overview of the current activities related to the surface component of IceCube.

\section{Current Analyses}
\label{sec:analysis}

A unique feature of IceCube is the measurement of the (muonic) shower nucleus with the in-ice instrumentation, where shower muons above 0.5$\,$TeV reach the in-ice detector and can be detected. The 3-dimensional structure of the in-ice Cherenkov light measured in the sensors is analyzed and used as a mass-sensitive parameter, in particular in correlation with the density of all charged particles of the shower at the surface measured with IceTop. 
The results contain a large uncertainty first due to the shower-to-shower fluctuations, and secondly due to shortcomings of the Monte Carlo simulations in the description of the EAS evolution, especially for the muons. It has been shown that a dedicated measurement of the muon densities at different energy thresholds can significantly improve the situation~\cite{Meurer:2005dt}. Therefore, analyses of the IceTop data focus on a possible identification of the shower muons. For that, at IceTop, a determination is made of the fraction of the measured detector signal produced by secondary muons. 
%
%
Here, nature, or better the shower development, helps in the way that the lateral distribution of the muons is flatter than that of the electrons, so that with larger distance to the shower core the relative signal fraction of muons becomes dominant. However, the total number of particles decreases quickly in the lateral direction, so that fewer particles are measured per shower.
It was also shown that the ratio of GeV muons to TeV muons is an important parameter for the cross-check of the hadronic interaction models~\cite{Engel:2019dsgf}. \\
For these reasons, recent analyses at IceTop have focused on determining the GeV muon density in detected air showers. A first analysis (fig.~\ref{fig2}) determines the mean muon density at large distances and correlates it with the energy of the primary particles~\cite{Soldin:2021icrc}. 
The measured densities are not coherently consistent with predicted muon densities obtained from the hadronic interaction models. \\
For a determination of the elemental composition, however, it is advantageous if the muon number or a muon density can be determined on the basis of individual events~\cite{Kang:2021icrc}. The paper suggests how promising this can be for determining the composition. \\  
The selection of muons in air showers improves when examining more horizontally incident primary particles, since here the electromagnetic components are already attenuated by the larger path length through the atmosphere. Therefore, the analysis of air showers up to 60 degrees zenith angle aims to check and verify the elemental composition of cosmic rays at IceCube with an independent set of events and thus to gain better determination of systematic uncertainties~\cite{Balagopal:2021icrc}. \\
The extracted muon density is also used in an analysis which focuses on the comparison, test and validity studies of hadronic interaction models~\cite{Verpoest:2021icrc}. IceCube has the capability to measure simultaneously the electromagnetic, GeV muon and TeV muon components of air showers. In that work, tests of various hadronic interaction models are presented by comparing data to proton and iron simulations for three different composition sensitive variables. If the models give a realistic description of experimental data, the composition interpretation of all variables should be consistent. However, IceTop indicates inconsistencies between different components, notably between the slope of the lateral distribution of the charged particles and the low-energy muons in all models. \\
In another analysis
air-shower signals seen in IceTop are used to determine a real-time veto for astronomical neutrino alerts sent out in order to trigger multi-messenger campaigns. From June 19, 2019 to December 31, 2020, IceCube sent 45 public alerts to the multi-messenger community. In this list, 6 alerts were for down-going ($\theta<82^\circ$) events including the four events with reconstructed energies above 1 PeV. IceTop data is used to tag cosmic ray induced event when there is a significant number of correlated IceTop pulses recorded with the in-ice muons within a time residual window of 0 to 1 $\mu$s. A count of at least 2 stations or 6 tanks of IceTop correlated in-time with the in-ice event marks IceTop activity. So far, using IceTop information, one alert event was cautioned as being of cosmic-ray origin and two high-energy events were retracted.    \\
Finally, improved analysis methods are applied to make more efficient use of the various air shower observables of the hybrid detector system IceCube. These are based in particular on machine learning and show promising preliminary results~\cite{Koundal:2021icrc}.

\section{Future Instrumentation}
\label{sec:future}

\subsection{Surface Array Enhancement}
\begin{figure}[t]
    \centering
    \includegraphics[height=6cm]{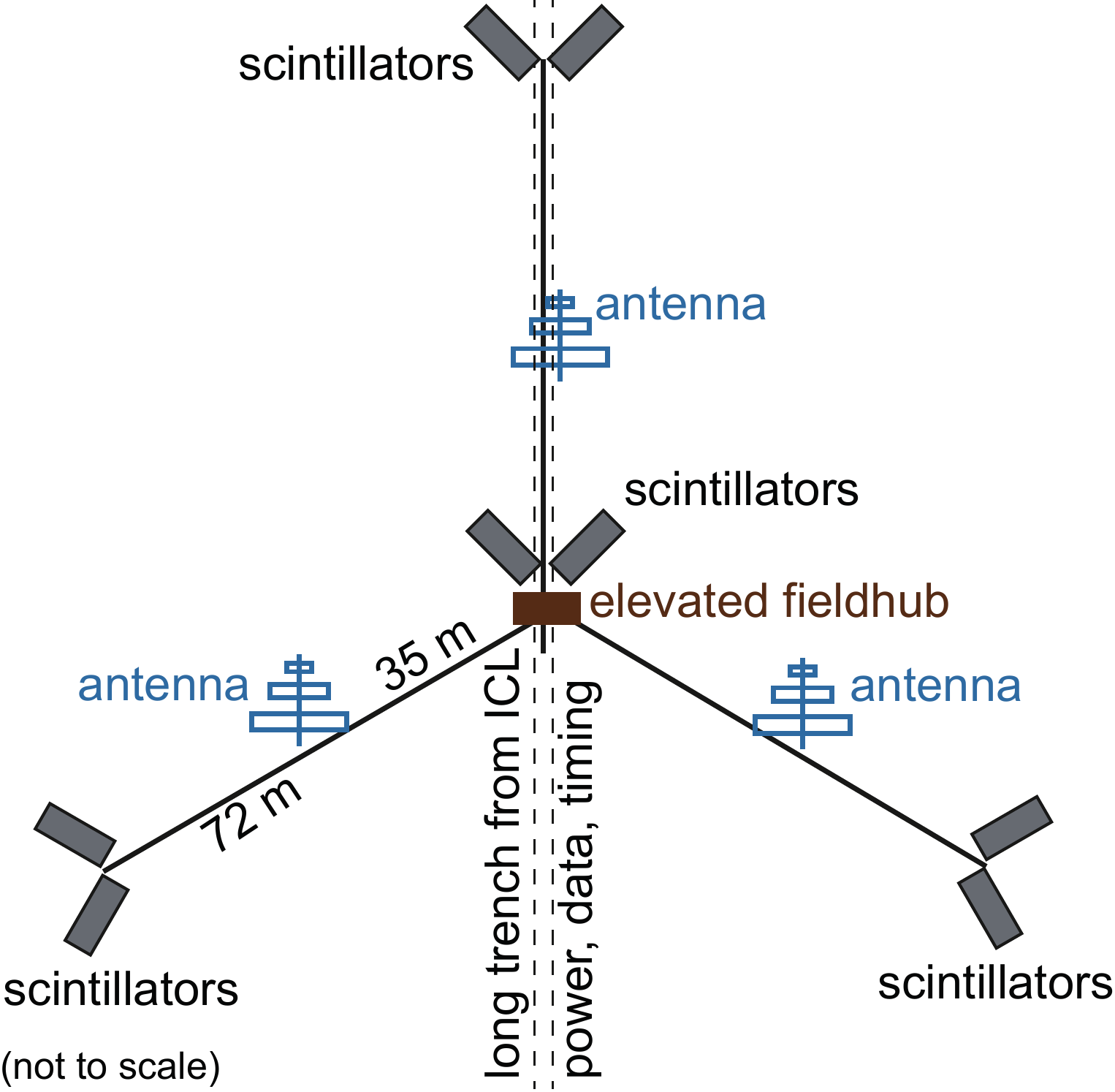}
    \hspace{1cm}
    \includegraphics[height=5cm]{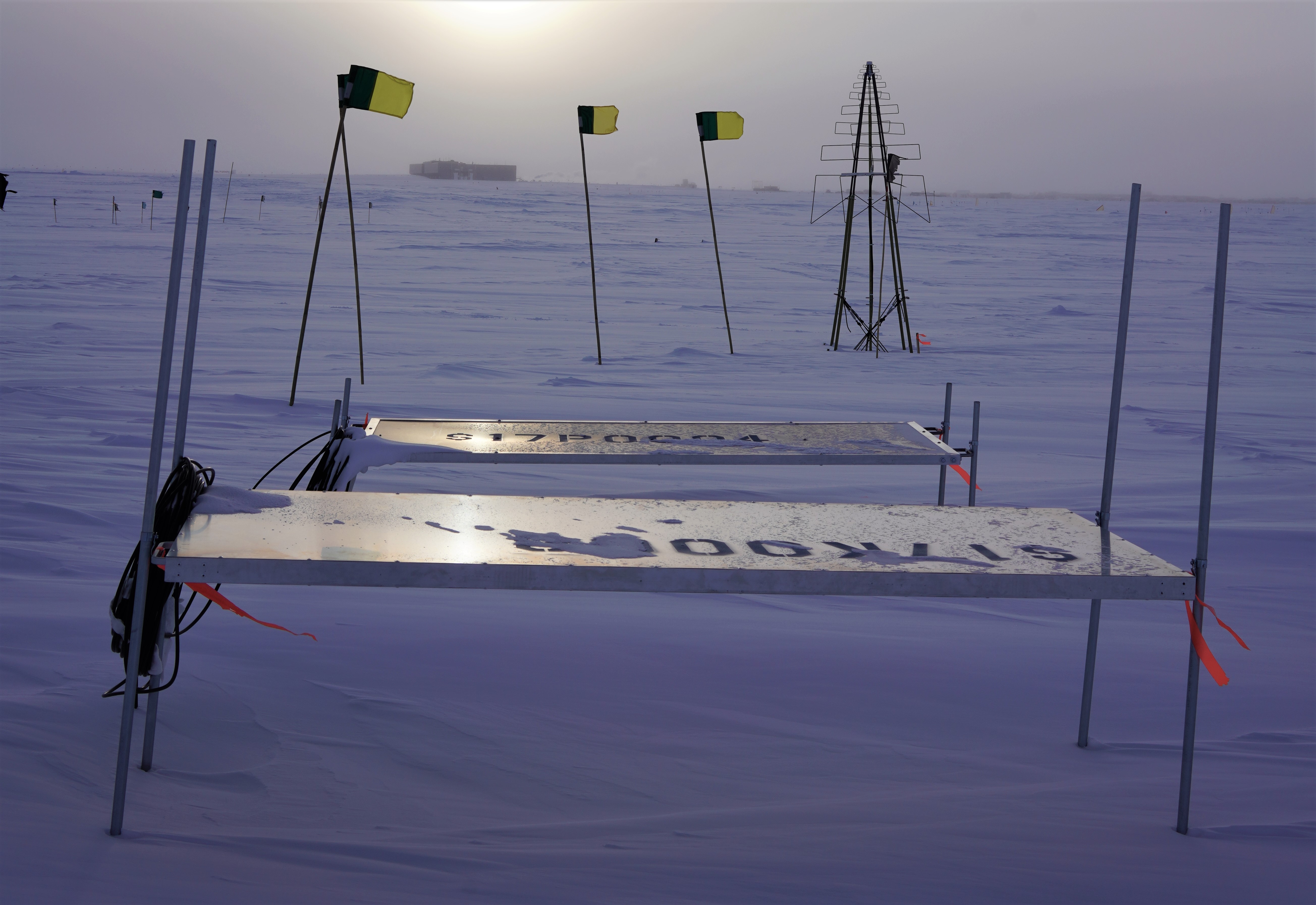}
    \caption{(Left) Layout of a surface station for the enhancement of IceTop, which is also the baseline design for the Gen2 surface array. (Right) Corresponding prototype detectors at IceTop; both the scintillators and radio antennas are deployed on stands that can be lifted to avoid snow management.}
    \label{fig:surface_layout}
\end{figure}

IceTop measures cosmic rays in the transition region from galactic to extra-galactic sources. However, the non-uniform snow accumulation on the installed ice-Cherenkov tanks leads to a non-uniform attenuation of the electromagnetic component which results in an increased uncertainty on the reconstruction of the air-shower parameters. 
Therefore, an upgrade of IceTop with an array of scintillator panels is under construction~\cite{Haungs:2019ylq}.
The enhancement foresees the deployment of 32 stations of 8 detectors and 3 radio antennas, read out by a central station DAQ, each within the present IceTop area~\cite{Oehler:2021icrc} (fig.~\ref{fig:surface_layout}). 
Taking advantage of the infrastructure that the scintillator array will provide, installation of radio antennas is also underway~\cite{Coleman:2021icrc}. 
These only moderate additional efforts will make the surface array to a multi-component detector. 
Furthermore, the collaboration has examined the possibility of adding Cherenkov telescopes (IceAct) to the surface instrumentation~\cite{Paul:2021icrc} that would measure the electromagnetic component of particularly lower energy air-showers as another complementary building block towards a hybrid cosmic-ray observatory.

The proposed detector types are optimized to serve the following general goals:
\begin{itemize}[itemsep=0pt,topsep=-1pt]
\item Cross-calibration: The coincident detection of air showers and muons deep in the ice will allow for an improved calibration of the in-ice detector and IceTop.
\item Improved capabilities for studying cosmic rays: The detection of comic rays through several independent detection channels will enhance the capabilities of In-ice IceCube and it's surface component IceTop to measure the mass composition of cosmic rays as well as allowing for composition dependent anisotropy studies.
\item Lowering the threshold for air-shower observations: The higher density of detectors will allow accurate reconstruction of air showers for energies below 1 PeV, i.e.\ the full energy range of the knee will be covered.
\item Better understanding of hadronic interaction models: The measurements of air showers through several detection channels will improve the understanding of hadronic interactions.
\item Improved sensitivity to primary gamma rays: For the gamma-ray detection from possible UHE sources a larger energy range and a larger exposure will be available.
\item Improvement of surface veto capabilities for atmospheric neutrinos: The energy threshold for vetoing the  background to astrophysical neutrinos in IceCube will be lowered. 
\end{itemize}

\begin{wrapfigure}{l}{0.45\textwidth}
  \begin{center}
    \includegraphics[width=0.43\textwidth]{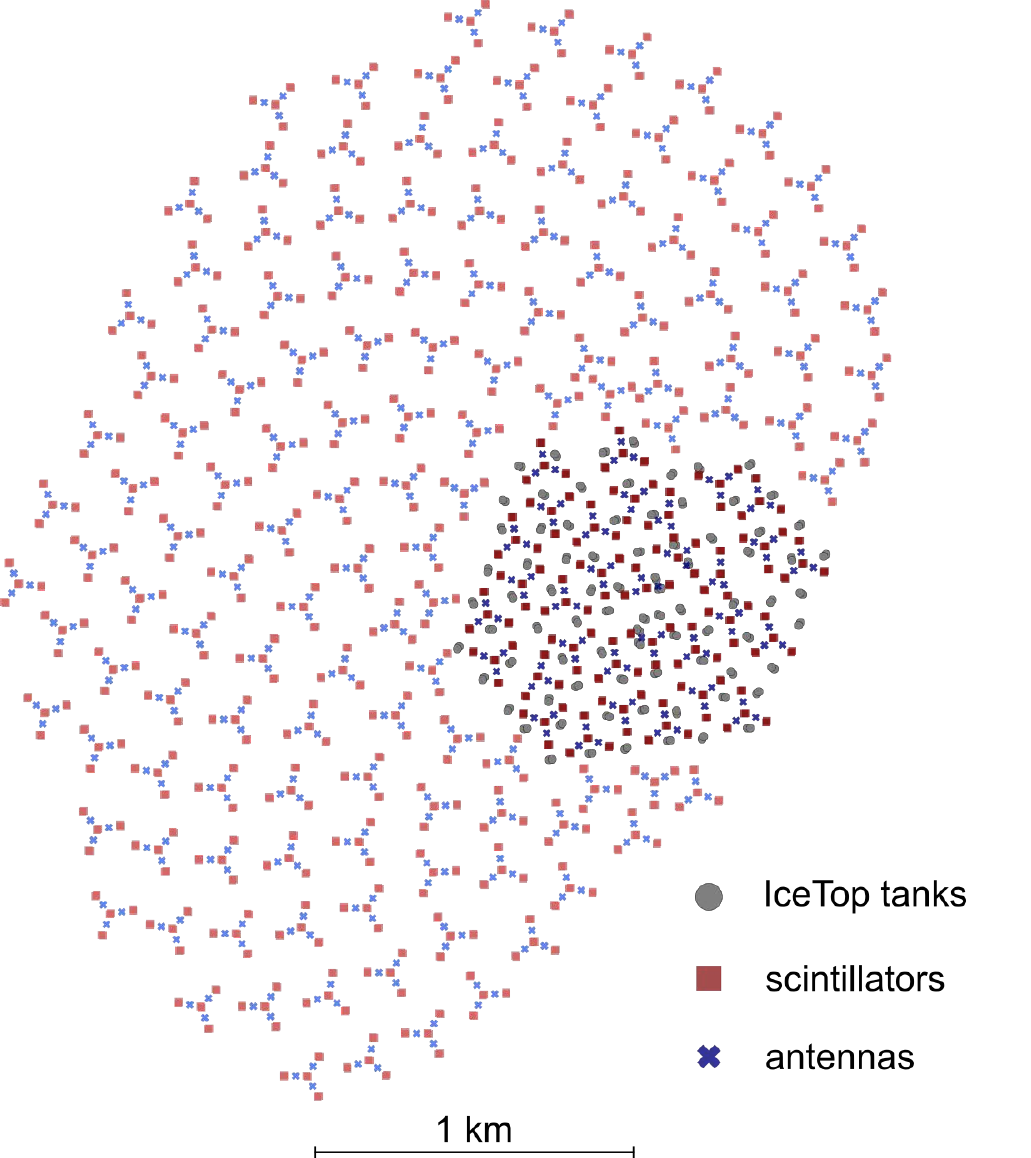}
  \end{center}
  \caption{Surface array of IceCube-Gen2 (darker colors) and IceTop enhancement (lighter colors) consisting of hybrid stations with eight scintillation detectors and three radio antennas. In addition IceTop tanks are shown.}
  \label{fig:gen2layout}
\end{wrapfigure}

Two R\&D scintillator stations with different designs were deployed in January 2018 and performed well. One of these stations was upgraded with two radio antennas in January 2019~\cite{Oehler:2021icrc}.
Using these experiences, a new prototype station combining and improving the previous iterations was designed.
This prototype station was deployed at the South Pole in January 2020, replacing the old stations.
With it, the first coincident measurements of cosmic-ray air-showers using scintillation detectors, radio antennas and IceTop were obtained~\cite{Dujmovic:2021icrc}.
Furthermore two prototype air-Cherenkov telescope are in operation at the South pole and have reported hybrid events detected with IceCube and IceTop~\cite{Paul:2021icrc}.
At this conference we present the hardware design and the performance of the prototype station as well as the plans for the full deployment.

\subsection{IceCube-Gen2 Surface Array}

The conceptual design of the surface instrumentation for IceCube-Gen2~\cite{Kowalski:2021icrc} will be similar to the enhancement planned for IceTop with correspondingly larger spacing. The baseline design assumes a station on top of each new in-ice string of IceCube-Gen2 (fig.~\ref{fig:gen2layout}).
With a spacing of $\sim$240\,m, such a surface array would provide hybrid measurements of the primary spectrum and mass composition from PeV to several EeV. 
A few additional stations between the current IceTop and the new surface array will guarantee a smooth coverage, enabling a consistent analysis of both surface arrays. 
Moreover, a small overlap with the Gen2 in-ice radio array will allow for the calibration of the cosmic-ray signals detected by the in-ice antennas. 
This overlap will ensure that all detector components of IceCube and Gen2 will share the same absolute energy scale by cross-calibration against the same air-shower array on the surface~\cite{Schroder:2021icrc}. 
\begin{figure}[t]
    \centering
    \includegraphics[width=0.9\textwidth]{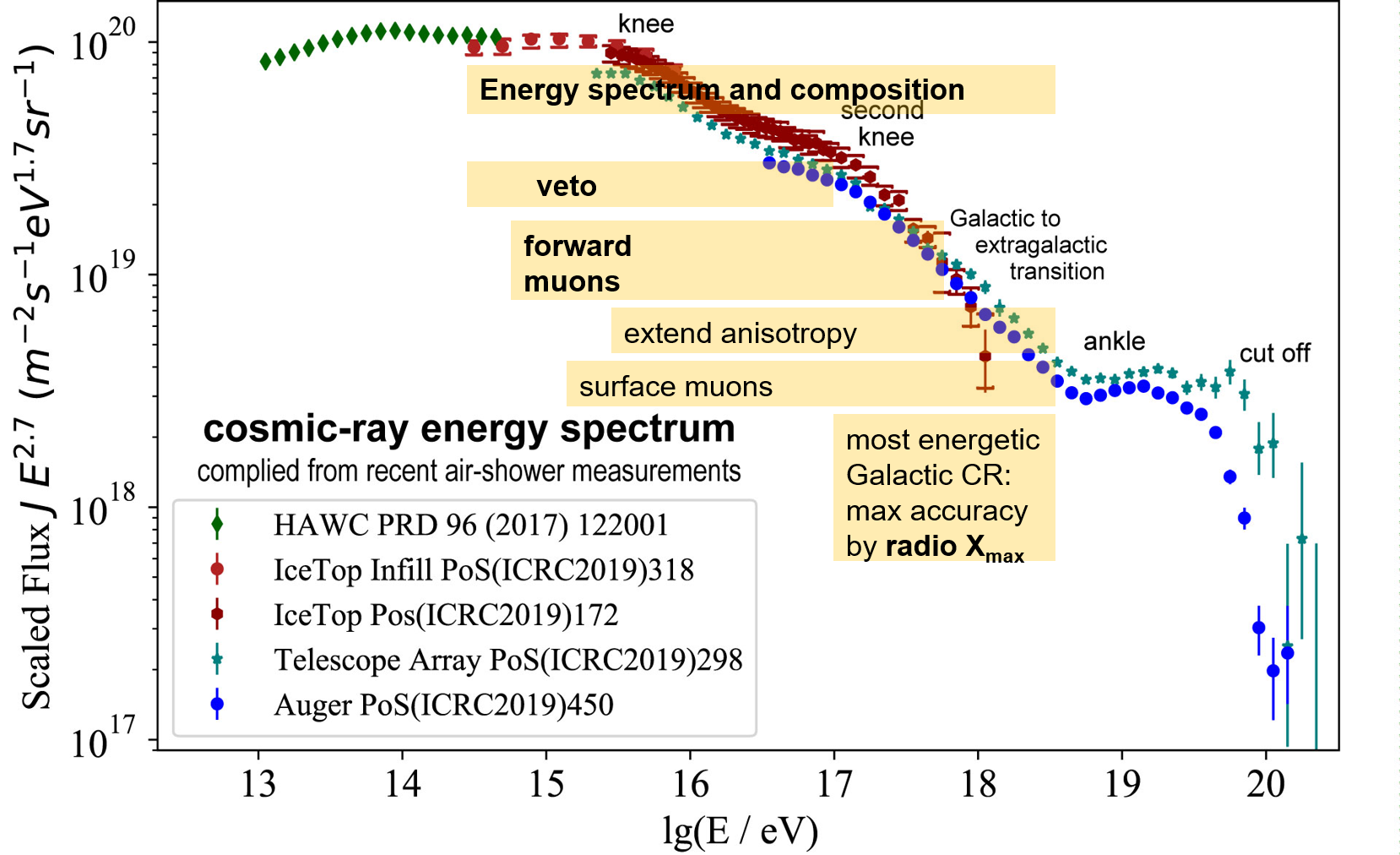}
    \caption{The energy spectrum of cosmic rays with indications (the yellow bands show the respective energy range) of the rich physics program of the IceCube and IceCube-Gen2 surface instrumentation~\cite{Schroder:2021icrc}.}
    \label{fig:spectrum}
\end{figure} 
%

With the larger area and the larger accessible angular range available, the acceptance for coincident measured surface and in-ice events increases by a factor of more than 30 compared to IceCube.
This leads to the unique possibility of an array with large acceptance for events in which the air shower at the surface and the bundle of $\sim$~TeV muons in the deep array are detected in coincidence.
The design will furthermore allow for selecting individual unaccompanied high-energy muons which can be used to calibrate the in-ice reconstruction of muons.
In addition, the TeV muons will provide information on the mass composition and on hadronic interactions in the air showers complementary to low-energy muons at the surface.
This is important to better understand the flux of atmospheric leptons creating background for the astrophysical neutrino measurements.
In addition, the drastically increased aperture for coincident events with the in-ice detectors increases the potential to directly discover nearby sources by PeV photons accordingly.
A surface detector also opens up the possibility of vetoing the background of cosmic-ray muon and even atmospheric neutrinos. 
With hundreds of coincident events per year above one EeV, such a detector would allow for an unprecedented measurement of the evolution of the primary composition in the region where a transition from Galactic to extra-galactic cosmic rays is predicted~\cite{Leszczynska:2021icrc}. 

\section{Summary}
\label{sec:summary}

IceCube with its surface array IceTop covers the complete range of high-energy GCRs from below 1~PeV to beyond 1~EeV. 
The simultaneous measurement of low-energy particles at the surface and high-energy muons in the ice offers unique opportunities for the study of hadronic interactions, and, in addition, also allows an improved search for PeV photons. 
A planned enhancement by a scintillator-radio hybrid array will significantly increase the accuracy and sky coverage of IceTop. 
Air-Cherenkov detectors can further enhance its accuracy around a few PeV and below. 
Finally, a planned expansion of IceCube to IceCube-Gen2 with a corresponding surface instrumentation will increase the exposure by an order of magnitude and will open a new window in studying the highest energy Galactic cosmic rays.

\bibliographystyle{ICRC}
\begin{small} 
\bibliography{ICRC2021_336_Haungs}
\end{small}












\end{document}